# The original Gibbs paradox is the consequence of the erroneous identification of non-identical functions

## Volodymyr Ihnatovych

Igor Sikorsky Kyiv Polytechnic Institute

e-mail: v.ihnatovych@kpi.ua

**Abstract**

This article presents the results of research into the causes of the Gibbs paradox in the formulation discussed by J. W. Gibbs himself. In this formulation, we are talking about an inexplicable (paradoxical) jump in the entropy of mixing of two ideal gases during the transition from mixing different to mixing identical gases.

It is shown that the entropy of mixing of different ideal gases and the entropy of mixing of identical ideal gases are different (non-identical) functions of the same gas parameters. That, called a paradoxical jump in the entropy of mixing, is not a jump in the value of some function, but is the difference in the values of various functions, on condition that the variables and parameters on which these functions depend remain constant. Those who were looking for an explanation of the original Gibbs paradox did not notice this and tried to solve an unsolvable falsely posed problem: to find a parameter that change during the transition from different to identical gases caused the difference in the values of non-identical functions.

## I. Introduction

The Gibbs Paradox is one of the most mysterious physical paradoxes. It has been known for over a hundred years. It was discussed by J. W. Gibbs, M. Planck, J. D. van der Waals, H. A. Lorentz, A. Sommerfeld, E. Schrödinger, P. W. Bridgman and other well-known physicists (see, e.g. [1–6]). Various explanations to this paradox have long been outlined in textbooks (see, e.g. [2,3,7–9]), but papers devoted to it appear again and again (see, e.g., [5,10–19]).

There are several formulations of the Gibbs paradox. The formulation, which we, following the example of J. van Lith ([16]), call the original Gibbs paradox, arises when considering the entropy of mixing of two ideal, chemically non-interacting gases with the same temperatures and pressures (which are initially separated by an impenetrable partition), in the framework of classical thermodynamics (see, for example, [1,2,4,6–19]).

J. W. Gibbs and others, by reasoning, concluded that the entropy of mixing of various ideal gases under the specified conditions is not equal to zero and does not depend on the properties of the gases being mixed. Also, by reasoning, the conclusion was obtained that the entropy of mixing of identical ideal gases is equal to zero. Thus, if we assume that there is a mixing of gases that are increasingly similar in properties and are with the same temperatures and pressures, then the entropy of mixing remains constant as long as there remains any, even vanishingly small, difference between the mixed gases. When a transition to identical ideal gases occurs, the entropy of mixing turns to zero by a jump, and the magnitude of the jump does not depend on what and how different these gases were. This behavior of the entropy of mixing is paradoxical: a quantity that does not depend on the degree of difference in properties of the gases suddenly vanishes when the difference in the properties disappears. This formulation is also called the thermodynamic version of the Gibbs paradox [13,17], the Gibbs paradox of the second kind [14], the discontinuity puzzle [18], and the first Gibbs paradox [19].

The jump in the entropy of mixing also looks paradoxical because the entropy of mixing, like the entropy of gases, is a certain function, i.e., a quantity that depends on other quantities (variables and parameters). The value of a function cannot change if the quantities on which it depends have not changed. Therefore, the change in the entropy of mixing during the transition from mixing different to mixing identical gases must be associated with a change in the value of some variable or parameter. In the case of the specified jump in the entropy of mixing, however, it is not clear which change in which parameter or variable causes this jump. Those who gave an explanation for the jump in the entropy of mixing associated it with jumps in various quantities, but there is still no explanation that would not cause objections, and which could be called final or generally accepted.

This situation has developed and persisted because the authors who discussed the original formulation of the Gibbs paradox did not take into account the fact that the conclusion about the paradoxical jump in the entropy of mixing is not based on the processing of empirical data, but is the conclusion of reasoning based on certain premises. They did not analyze this reasoning and did not notice that the indicated conclusion is based on a comparison of the values of non-identical functions, which are erroneously considered to be the same. To demonstrate this, we derive a series of equations and consider the question: why do the values of the entropy of mixing of different and identical ideal gases differ?

## II. Derivation of equations for the entropy of mixing of ideal gases, initially separated by a partition, and obtaining a conclusion about a paradoxical jump

When deriving equations for the entropy of mixing of different and identical ideal gases, we will use as initial equations the well-known equations for the entropy of an ideal gas, the entropy of systems of ideal gases, the change in entropy with a change in the state of a thermodynamic system, and the equation of state of an ideal gas:

$$S_i(n_i, V_i, T_i) = n_i \left( c_{Vi} \ln T_i + R \ln \frac{V_i}{n_i} + s_{0i} \right), \qquad (1)$$

$$S_b(n_1, V_1, T_1, n_2, V_2, T_2) = S_1(n_1, V_1, T_1) + S_2(n_2, V_2, T_2), \qquad (2)$$

$$S_m(n_1, n_2, V, T) = S_1(n_1, V, T) + S_2(n_2, V, T), \qquad (3)$$

$$\Delta S = S_{II} - S_I, \qquad (4)$$

$$p_i V_i = n_i R T_i, \qquad (5)$$

where $S_i(n_i, V_i, T_i)$ is the entropy of $n_i$ moles of the $i$th ideal gas, its volume is $V_i$ and temperature is $T_i$; $c_{Vi}$ is the molar heat capacity of the $i$th ideal gas at a constant volume, which depends on the nature of the gas; $R$ is universal gas constant; $s_{0i}$ is a constant, which depends on the nature of the gas; $S_b(n_1, V_1, T_1, n_2, V_2, T_2)$ is the entropy of a system consisting of two ideal gases separated by an impermeable partition, the quantities of which are $n_1$ and $n_2$ moles, the volumes are $V_1$ and $V_2$, and the temperatures are $T_1$ and $T_2$; $S_m(n_1, n_2, V, T)$ is the entropy of a mixture of $n_1$ and $n_2$ moles of different ideal gases, its volume is $V$ and temperature is $T$; $\Delta S$ is change in the entropy of the system during its transition from initial state I to final state II; $S_I$ is the entropy of the system in the initial state, $S_{II}$ is the entropy of the system in the final state; $p_i$ is the pressure of the $i$th ideal gas.

The Eq. (3) expresses the Gibbs' theorem on the entropy of a mixture of ideal gases (see e.g. [7,8,10]).

The Eqs. (1)–(5) or equivalent equations are used by other authors who considered the Gibbs paradox (see, e.g. [1–3,7–10,12,16]).

Function designations in Eqs. (1)–(3) contain variables on which these functions depend. In what follows, the designations of various functions will also indicate the variables on which they depend. The values related to systems of different gases are denoted by superscripts *d*, and the values related to systems of identical ideal gases will be denoted by superscripts *i*.

J. W. Gibbs considered the case of mixing ideal gases with the same temperatures, pressures and volumes [1]. We will consider a more general case — the mixing of ideal gases with the same initial temperatures, but different volumes and pressures. Since the gases are ideal, the temperature of the system after mixing will be equal to the temperature of the gases before mixing.

Suppose that $n_1$ and $n_2$ moles of different ideal gases 1 and 2, with identical temperatures equal to $T$, and different volumes $V_1$ and $V_2$, are separated by an impermeable partition. The next equation for the entropy of this system follows from Eqs. (1) and (2):

$$S_b^d(n_1, V_1, n_2, V_2, T) = (n_1 c_{V1} + n_2 c_{V2}) \ln T + R\left(n_1 \ln \frac{V_1}{n_1} + n_2 \ln \frac{V_2}{n_2}\right) + n_1 s_{01} + n_2 s_{02}. \quad (6)$$

After removing the partition, a mixture of $n_1$ and $n_2$ moles of ideal gases 1 and 2 of volume $V_1 + V_2$ is formed. The next equation for the entropy of this mixture follows from Eqs. (1) and (3):

$$S_m(n_1, n_2, V_1 + V_2, T) = (n_1 c_{V1} + n_2 c_{V2}) \ln T + R\left(n_1 \ln \frac{V_1 + V_2}{n_1} + n_2 \ln \frac{V_1 + V_2}{n_2}\right) + n_1 s_{01} + n_2 s_{02}. \quad (7)$$

The next equation for the entropy of mixing of different ideal gases 1 and 2 with equal initial temperatures follows from Eqs. (4), (6), (7), taking into account the fact that $S_I^d = S_b^d(n_1, V_1, n_2, V_2, T)$ and $S_{II}^d = S_m(n_1, n_2, V_1 + V_2, T)$:

$$\Delta S^d(n_1, V_1, n_2, V_2) = R\left(n_1 \ln \frac{V_1 + V_2}{n_1} + n_2 \ln \frac{V_1 + V_2}{n_2}\right) - R\left(n_1 \ln \frac{V_1}{n_1} + n_2 \ln \frac{V_2}{n_2}\right). \quad (8)$$

We transform Eq. (8):

$$\Delta S^d(n_1, V_1, n_2, V_2) = R(n_1 + n_2) \ln \frac{V_1 + V_2}{n_1 + n_2} - R\left(n_1 \ln \frac{V_1}{n_1} + n_2 \ln \frac{V_2}{n_2}\right) + \\ + R[(n_1 + n_2) \ln(n_1 + n_2) - (n_1 \ln n_1 + n_2 \ln n_2)] \quad (9)$$

If ideal gases have the same temperature and pressure, then it follows from Eq. (5):

$$\frac{V_1}{n_1} = \frac{V_2}{n_2} = \frac{V_1 + V_2}{n_1 + n_2}. \quad (10)$$

The next equation for the entropy of mixing of different ideal gases with the same temperatures and pressures, separated before mixing by an impermeable partition, follows from (9) and (10):

$$\Delta S^d(n_1,n_2) = R[(n_1+n_2)\ln(n_1+n_2) - (n_1 \ln n_1 + n_2 \ln n_2)]. \tag{11}$$

If also $n_1 = n_2 = n$, then from (11) it follows:

$$\Delta S^d(n_1,n_2) = 2Rn\ln 2. \tag{12}$$

If also $n_1 = n_2 = 1$, then from (11) it follows:

$$\Delta S^d(n_1,n_2) = 2R\ln 2. \tag{13}$$

Now let us consider the case of mixing of identical ideal gases with the same temperatures.

Suppose, $n_1$ and $n_2$ moles of identical gases 3 and 3, with temperatures equal to $T$, and volumes of $V_1$ and $V_2$, are separated by an impenetrable partition. The next equation for the entropy of this system follows from Eqs. (1) and (2):

$$S_b^i(n_1,V_1,n_2,V_2) = (n_1+n_2)c_{V3} \ln T + R\left(n_1 \ln \frac{V_1}{n_1} + n_2 \ln \frac{V_2}{n_2}\right) + (n_1+n_2)s_{03}. \tag{14}$$

After removing the partition, $n_1 + n_2$ moles of pure ideal gas 3 of volume $V_1 + V_2$ is formed, that entropy according to Eq. (1) is equal:

$$S_3(n_1+n_2, V_1+V_2, T) = (n_1+n_2)c_{V3} \ln T + R(n_1+n_2)\ln \frac{V_1+V_2}{n_1+n_2} + (n_1+n_2)s_{03}. \tag{15}$$

The next equation for the entropy of mixing of identical ideal gases with equal initial temperatures follows from Eqs. (4), (14), (15), taking into account the fact that $S_I^i = S_b^i(n_1,V_1,n_2,V_2,T)$ and $S_{II}^i = S_3(n_1+n_2, V_1+V_2, T)$:

$$\Delta S^i(n_1,V_1,n_2,V_2) = R(n_1+n_2)\ln\frac{V_1+V_2}{n_1+n_2} - R\left(n_1 \ln \frac{V_1}{n_1} + n_2 \ln \frac{V_2}{n_2}\right). \tag{16}$$

The next equation for the entropy of mixing of identical ideal gases with the same temperatures and pressures, separated before mixing by an impermeable partition, follows from (16) and (10):

$$\Delta S^i(n_1,n_2) = 0. \tag{17}$$

Based on Eqs. (11)–(13) and (17), we obtain a conclusion about the paradoxical behavior of the entropy of mixing in the transition from mixing different to mixing identical gases. The entropy of mixing of different ideal gases with the same temperatures and pressures $\Delta S^d(n_1,n_2)$, according to Eqs. (11)–(13), is not equal to zero, it depends only on the quantities of gases $n_1$ and $n_2$ and does not depend on the gases' properties. The entropy of mixing of identical ideal gases with the same temperatures and pressures $\Delta S^i(n_1,n_2)$, according to Eq. (17), is equal to zero. Let us assume that, at constant values of $n_1$ and $n_2$ the properties of gases 1 and 2 approach the properties of gas 3 and eventually, gases 1 and 2 become identical to gas 3. The entropy of mixing, according to Eqs. (11)–(13), remains constant as long as gases 1 and 2 are different. When gases 1 and 2 become identical to gas 3, their entropy of mixing, according to Eq. (17), is equal to zero. Thus, during the transition from mixing different ideal gases to mixing identical ideal gases, the entropy of mixing turns to zero by a jump.

In the same way, by comparing the values of $\Delta S^i(n_1, n_2)$ and $\Delta S^d(n_1, n_2)$ for cases of mixing different and identical ideal gases with the same temperatures and pressures, other authors also obtained a conclusion on the paradoxical jump in the entropy of mixing (see, e.g. [1,2,7–10,12,16]). The conclusion about the paradoxical jump in the entropy of mixing seems absolutely reliable. This conclusion follows from a comparison of Eq. (17) with Eqs. (11)–(13). Eqs. (11)–(13), (17) are obtained on the basis of generally accepted initial equations by mathematical derivations given above. The correctness of the derivation of each of these equations can be easily verified, since all the initial equations and all the assumptions used in obtaining each equation are given above, and when obtaining each intermediate equation, it is indicated from which initial and intermediate equations it follows.

### III. Identification of the error, the consequence of which is the conclusion about the paradoxical jump in the entropy of mixing

Consider the question: what causes the difference in the values of the entropy of mixing of the same quantities of different and identical ideal gases with the same initial temperatures, i.e., the difference in the values of the functions $\Delta S^d(n_1, V_1, n_2, V_2)$ and $\Delta S^i(n_1, V_1, n_2, V_2)$, expressed by Eqs. (9) and (16) or Eqs. (11) and (17), which are special cases of Eqs. (9) and (16)?

To find the answer to this question, we obtain an equation relating the quantities $\Delta S^d(n_1, V_1, n_2, V_2)$ and $\Delta S^i(n_1, V_1, n_2, V_2)$. It follows from (9) and (16) or (11) and (17):

$$\Delta S^d(n_1, V_1, n_2, V_2) - \Delta S^i(n_1, V_1, n_2, V_2) = R[(n_1 + n_2)\ln(n_1 + n_2) - (n_1 \ln n_1 + n_2 \ln n_2)] \quad (18)$$

It follows from (18):

$$\Delta S^d(n_1, V_1, n_2, V_2) = \Delta S^i(n_1, V_1, n_2, V_2) + R[(n_1 + n_2)\ln(n_1 + n_2) - (n_1 \ln n_1 + n_2 \ln n_2)] \quad (19)$$

From the Eq. (19) it can be seen that the function $\Delta S^d(n_1, V_1, n_2, V_2)$, called the entropy of mixing of different ideal gases, is equal to the sum of two functions: the function $\Delta S^i(n_1, V_1, n_2, V_2)$, called the entropy of mixing of identical ideal gases, and the function $R[(n_1 + n_2)\ln(n_1 + n_2) - (n_1 \ln n_1 + n_2 \ln n_2)]$. Therefore, the function $\Delta S^d(n_1, V_1, n_2, V_2)$, called the entropy of mixing of different ideal gases, is not identical to the function $\Delta S^i(n_1, V_1, n_2, V_2)$, called the entropy of mixing of identical ideal gases.

The value $R[(n_1 + n_2)\ln(n_1 + n_2) - (n_1 \ln n_1 + n_2 \ln n_2)]$, which is interpreted as the magnitude of the jump in the entropy of mixing ideal gases during the transition from mixing different gases to mixing identical gases, is not a change in the value of some function, but is the difference in the values of non-identical functions $\Delta S^d(n_1, V_1, n_2, V_2)$ and $\Delta S^i(n_1, V_1, n_2, V_2)$ provided that all variables and parameters on which these functions depend, namely $n_1$, $n_2$, $V_1$, $V_2$, have the same values. Precisely because the function $\Delta S^d(n_1, V_1, n_2, V_2)$, which is expressed by Eqs. (9) or (11), is not identical to the function $\Delta S^i(n_1, V_1, n_2, V_2)$, which is expressed by Eqs. (16) or (17), the value of the so-called entropy of mixing of different gases differs from the value of the so-called entropy of mixing of identical gases.

However, J. W. Gibbs and other authors, who made a conclusion about the paradoxical jump in the entropy of mixing, erroneously believed that Eqs. (11) and (17) do not express different functions, but different values of the same function — the entropy of mixing of ideal gases. The value of a function, however, cannot change if the values of the parameters and variables on which it depends do not change. Therefore, those who were looking for an explanation for the paradoxical jump in the entropy of mixing tried to find a parameter that changes abruptly in the transition from mixing different to mixing identical gases, or to connect the indicated jump with some difference in the mixing of different and identical gases.

J. W. Gibbs associated the difference in the values of the entropy of mixing of different and identical gases with the fact that "the mixture of gas-masses of the same kind stands of a different footing from the mixture of gas-masses of different kind" [1]. M. Planck explained the jump in the entropy of mixing by the fact "that the chemical difference of two gases… cannot be represented by a continuous variable" [2, p. 237]. I. P. Bazarov associated the jump in the entropy of mixing with the jump in the change in the density of mixed gases during the transition from mixing different to mixing identical gases [7]. M. W. Zemansky and R. H. Dittman argued that P. W. Bridgman resolved the Gibbs paradox. This explanation is based on the argument that a series of experimental operations are required to distinguish between two gases, but there is no instrumental operation by which identical gases could be distinguished; this causes a discontinuity of the function of the entropy change [8]. According to D. Dieks, the jump in the entropy of mixing during the transition to identical gases is due to the fact that two gases can either be separated or not [17].

These and other authors, who made the conclusion about the paradoxical jump in the entropy of mixing during the transition from mixing different to mixing identical gases, overlooked the fact that this conclusion is obtained on the basis of a comparison of certain equations, in this article these are Eqs. (17) and (11). If based on a comparison of Eq. (17) with Eq. (11), a conclusion is made about a jump in the entropy of mixing upon transition from different to identical gases, then it should be concluded that the entropy of mixing changes from $R[(n_1 + n_2)\ln(n_1 + n_2) - (n_1 \ln n_1 + n_2 \ln n_2)]$ to zero.

Accordingly, if we proceed from the fact that the jump in the entropy of mixing is due to a jump in some gas parameter, then it is necessary to explain how the jump in this parameter leads to the fact that the value of the function $R[(n_1 + n_2)\ln(n_1 + n_2) - (n_1 \ln n_1 + n_2 \ln n_2)]$ becomes equal to zero, under the condition that $n_1$ and $n_2$ keep constant values. Of course, this problem is insoluble. Therefore, no one could solve it for more than a hundred years. But this problem is false — it arises due to the erroneous identification of non-identical functions $\Delta S^d(n_1, V_1, n_2, V_2)$ and $\Delta S^i(n_1, V_1, n_2, V_2)$ and the erroneous interpretation of the function $R[(n_1 + n_2)\ln(n_1 + n_2) - (n_1 \ln n_1 + n_2 \ln n_2)]$ as the magnitude of the change in a certain function — the entropy of mixing.

One can ask the question: why did many physicists throughout the years not notice an obvious error and accepted that the functions $\Delta S^d(n_1, V_1, n_2, V_2)$ and $\Delta S^i(n_1, V_1, n_2, V_2)$ are identical, although the first is the sum of the second and the function $R[(n_1 + n_2)\ln(n_1 + n_2) - (n_1 \ln n_1 + n_2 \ln n_2)]$? Apparently the main reason for this is that they considered the conclusion about the paradoxical jump in the entropy of mixing as an incomprehensible (strange, paradoxical) fact. They did not take into account the fact that this paradoxical conclusion is the result of certain reasoning and it concerns the behavior of a certain function of many variables, which is expressed by equations. They did not attempt to analyze the reasoning that leads to the paradox, and when discussing the cause of the jump in the entropy of mixing, they did not take into account the equations that express this function. Without this, it is impossible to find that the conclusion about the paradoxical jump is made on the basis of comparing the values of non-identical functions.

## IV. Conclusions

The appearance of original Gibbs paradox in the formulation of the conclusion on a paradoxical jump in the entropy of mixing of ideal gases during the transition from mixing of different gases to mixing of identical gases is due to the erroneous identification of non-identical functions: the entropy of mixing of different ideal gases and the entropy of mixing of identical ideal gases, the first of which is the sum of the second and the function $R[(n_1 + n_2)\ln(n_1 + n_2) - (n_1 \ln n_1 + n_2 \ln n_2)]$. This error results in a completely unsolvable problem: to find a parameter of an ideal gas different from $n_1$ and $n_2$, which changes when passing from different gases to identical ones and causing the function $R[(n_1 + n_2)\ln(n_1 + n_2) - (n_1 \ln n_1 + n_2 \ln n_2)]$ to become zero.

If this error is not made and it is recognized that the function called the entropy of mixing of different ideal gases is not identical to the function called the entropy of mixing of identical ideal gases, then the difference in their values will become impossible to interpret as a change (jump) in the value of some function, and the original Gibbs paradox will not arise.

## Acknowledgements

The author is grateful to professor Oleksandr Andriiko, senior lecturer Vasiliy Pikhorovich and the late associate professor Viktor Haidey for helpful discussions and comments.## References

[1] Gibbs J W 1928 *The Collected Works Vol 1 Thermodynamics* (New York: Longmans)

[2] Planck M 1903 *Treatise on Thermodynamics* (New York: Longmans)

[3] Sommerfeld A 1956 *Thermodynamics and Statistical Mechanics* (New York: Academic Press)

[4] Khaytun S D 2010 *The History of the Gibbs Paradox* 3th edition (Moscow: KomKniga) (in Russian)

[5] *Gibbs paradox and its resolutions* / Compiled by S-K Lin [Internet]; 2009 Nov 7 [cited 2022 Dec 12]; [about 1 screen]. Available from: http://www.mdpi.org/lin/entropy/gibbs-paradox.htm

[6] Darrigol O 2018 The Gibbs Paradox: Early History and Solutions *Entropy* **20**, 443

[7] Bazarov I P 1991 *Thermodynamics* 4th edition, rev and enl (Moscow: Vysshaya Shkola) (in Russian)9